\documentclass{aa} 
\usepackage{graphicx}
\begin{document}
\title{Planetary companion candidates around the K giant stars 42~Dra and HD~139357}


   \author{
	M.P. D\"{o}llinger\inst{1}
          \and
	A.P. Hatzes\inst{2}
          \and
	L. Pasquini\inst{1}
          \and
   E.W. Guenther\inst{2}
          \and
	M.~Hartmann\inst{2}
          \and
	L. Girardi\inst{3}
             }

   \offprints{
    Michaela P. D\"{o}llinger \email{mdoellin@eso.org.de}\\$*$~
    Based on observations obtained at the 
    2m Alfred Jensch telescope at the Th\"uringer
    Landessternwarte Tautenburg 
  \\}

     \institute{
European Southern Observatory, Karl-Schwarzschild-Strasse 2, D-85748,
Garching bei M\"{u}nchen, Germany
         \and
	Th\"uringer Landessternwarte Tautenburg,
                Sternwarte 5, D-07778 Tautenburg, Germany
		\and 
INAF-Osservatorio Astronomico di Padova, Vicolo dell'Osservatorio 5, I-35122
Padova, Italy}

   \date{Received; accepted}

\abstract
{
For the past 3 years we have been monitoring 62 K giant stars
using precise stellar radial velocity (RV) measurements
with the 2m Alfred Jensch Telescope of the Th\"uringer Landessternwarte
Tautenburg ($TLS$).}
{To probe the dependence of planet formation on stellar mass
by finding planets around intermediate-mass giant stars.}
{We present high accuracy RV measurements of the K1.5~III
star 42~Dra and the K4~III star HD~139357. The wavelength reference for the
RV measurements was provided by an iodine absorption cell placed in the
optical path of the spectrograph.}
   {
Our measurements reveal that the time series of the radial velocity of
42~Dra shows a periodic variation of 479.1~days with a semiamplitude of
$K$~=~112.5~m\,s$^{-1}$. An orbital solution yields a mass function of
$f(m)$ $=$ (5.29 $\pm$ 0.62) $\times$ 10$^{-8}$ solar masses (M$_{\odot}$)
and an eccentricity of $e$~$=$~0.38~$\pm$~0.06.
From our template spectra, taken without the iodine cell, we determine a
metallicity of $-$0.46 $\pm$ 0.04 dex and a stellar mass of
0.98 $\pm$ 0.06 M$_{\odot}$ for this star.\\
HD 139357 shows periodic RV variations of
1125.7 days with a semiamplitude $K$~=~159.9~m\,s$^{-1}$. An orbital solution
yields an eccentricity, $e$ = 0.10 $\pm$ 0.02 and mass
function, $f(m)$ = (4.79 $\pm$ 0.57) $\times$ 10$^{-7}$ M$_{\odot}$.
An iron abundance of $-$0.13 $\pm$ 0.04 dex is obtained, and a stellar mass
of 1.31 $\pm$ 0.24 M$_{\odot}$ for the parent star is derived.
An analysis of the $HIPPARCOS$ photometry as well as our H$\alpha$
core flux measurements reveal no variability with the radial velocity
period. Keplerian motion is the most likely explanation for the
observed radial velocity variations for these stars.
}
   {
The K giant stars 42 Dra and HD 139357 host extrasolar planets with
``minimum masses'' of 3.88 $\pm$ 0.85 Jupiter masses M$_{\mathrm{Jup}}$ and
9.76~$\pm$~2.15~M$_{\mathrm{Jup}}$, respectively.
}

\keywords{star: general - stars: variable - stars: individual:
    \object{42~Dra, HD 139357}  - techniques: radial velocities -
stars: late-type - planetary systems}
\titlerunning{Planet candidates around 42 Dra and HD 139357}
\maketitle

%
\section{Introduction}

Until now more than 250 extrasolar planets around solar-type main-sequence
(MS) host stars have been detected via the RV method.
Out of these, the number of discovered planetary companions around giant stars
is still limited.\\
Our motivation was to enlarge the regime of stellar masses and characteristics
of planet host stars by hunting for planets around K giants to
provide us with important clues to the process of planet formation in a still
poorly investigated star domain. 
This is possible because giant host stars tend to be more massive than the 
solar-mass objects that have been the traditional targets of most planet
searches, thus we can probe the dependence of stellar mass on planet
formation. In addition these stars are in a different evolutionary status
and have different radii and internal structures compared to their
MS counterparts. Finally, in contrast to MS stars, the
chemical composition of planet host giants shows no preference for
metal-rich systems, which has been interpreted as an evidence for MS
pollution (Pasquini et al. 2007).\\
Hatzes $\&$ Cochran (1993) found first indications of substellar companions 
around giants. They discovered long-period RV variations in three
K giants and they proposed two viable hypotheses for these variations
in radial velocity: substellar companions or rotational modulation.
Recently Hatzes et al. (2006) and Reffert at al. (2006) confirmed 
that the initial RV variations found by Hatzes $\&$ Cochran (1993) 
in $\beta$ Gem were in fact due to a planetary companion.  
Since the first unequivocal discovery of the first extrasolar planet 
around the K giant HD 137759 ($\iota$~Dra) by Frink et al. (2002), several
giant stars have been found to host giant planets
(Setiawan et al. 2003a, 2003b, 2005; Sato al. 2003; Hatzes et al. 2005).
Currently, a number of groups are actively searching for extrasolar planets
around giant stars and this has resulted in a burst of recent discoveries
(Sato et al. (2007); D\"{o}llinger et al. (2007); Niedzielski 
et al. (2007); Johnson et al. (2007b); Sato et al. (2008).

\section{Observations and data analysis}
42 Dra and HD 139357 belong to a star sample observed since 
February 2004 from the Th\"uringer Landessternwarte Tautenburg ($TLS$) as part 
of the Tautenburg Observatory Planet Search Programme ($TOPS$). This programme
uses the coud{\'e} {\'e}chelle spectrograph
mounted on the 2m Alfred Jensch telescope.
This instrument provides a resolving power of $R$~=~67{,}000 and a wavelength 
coverage of 4700--7400~Angstr\"{o}ms~({\AA}s) using the so-called 
"visual" (VIS) grism mode.
The exposure time for both stars ranged between 5--10~minutes depending 
on the weather conditions and this resulted in a signal-to-noise (S/N) ratio 
typically greater than 150. Standard CCD data reduction
(bias-subtraction, flat-fielding and spectral extraction) was performed
using standard $IRAF$ routines. An iodine absorption cell placed
in the optical path provided the wavelength reference for the velocity
measurements.\\  

The RVs were computed using a programme that largely follows the 
prescription of Butler et al. (1996). It models an observation
of the star observed through the cell using a high resolution template
spectrum of molecular iodine taken with a Fourier Transform Spectrometer
($R$~$\approx$~10$^6$) and a template spectrum of the star taken without
the iodine cell. The programme takes
into account possible changes of the instrumental profile (IP) of the 
spectrograph by using the procedure outlined in Valenti et al. (1995).
Since the IP can also change spatially along a spectral order, these 
were divided into segments (so-called spectral chunks). About 
130~chunks were used in the final analysis. Thus the spatial (and temporal) 
variations of the IP, which can introduce significant RV errors, were modeled
independently for each chunk. Relative wavelength shifts between the iodine
and stellar template spectra were computed using a version of the
Fahlman $\&$ Glaspey (1973) shift-detection algorithm. 
The RV measurements from all chunks were then combined
weighted by the inverse square of the RV standard deviation for each chunk.
The internal velocity error of a spectrum is thus the dispersion of all
segments used for the analysis. We should note that the measured Doppler
shifts are relative to the stellar template and are thus
not absolute radial velocities.
For the bright K giant stars in our programme we typically achieve
a RV precision of about 3--5~m\,s$^{-1}$. This value refers to the internal 
velocity error.\\

To exclude rotational modulation as the cause of the RV variations in the
42 Dra and HD 139357 data, we investigated stellar activity
in these stars via the H$\alpha$ spectral line.
Several studies have shown that the central core of H$\alpha$ forms in the
chromosphere. Consequently stars of different chromospheric
activity show a different shape (depth and breadth) of the H$\alpha$ core
(Pasquini $\&$ Pallavicini 1991).
It has been widely demonstrated that the H$\alpha$ line can be used as
a good ``activity indicator'' of chromospheric emission (Herbig 1985;
Pasquini $\&$ Pallavicini 1991; Freire Ferrero et al. 2004).
By measuring the variations of the core of
H$\alpha$ with respect to the continuum it is therefore possible
to investigate the presence and variability of chromospheric active regions.

To determine the chromospheric contribution to the H$\alpha$ profile one
has to exclude possible contamination from telluric lines.
The normal procedure is to observe 
so-called ``telluric standards'' in order to divide out the 
telluric lines. These are typically early-type, 
rapidly rotating stars which are observed every night. 
However, with the spectra of the Tautenburg star sample this subtraction
was not done in this way because early-type stars have very strong balmer
lines so using them for taking out telluric lines is not so appropriate.
We therefore restricted our H$\alpha$ measurements to the
region within $\pm$~0.6~{\AA} of the line center. Two additional
spectral regions located at $\pm$~50~{\AA} provided the ``continuum''
measurement. All of these regions were free of telluric
contamination. The H$\alpha$ core measurement was done using a programme
that shifts all the spectra to a common rest wavelength and measures
the flux ratio of the defined areas (Biazzo 2007, private communication).
After running the programme, 74 Dra shows a variability of around 1~$\%$ in
H$\alpha$ activity. Taking into account that this star was not observed 
every night in all observing runs this value can only give a hint to the 
internal error of the measurements. Due to the H$\alpha$ activity of only 
around 1~$\%$ and RV variations at a very low level, which corresponds to 
the RV precision limit of about 3--5~m\,s$^{-1}$, 74 Dra was used as an 
internal H$\alpha$ and RV standard. The RV plot of this star is shown in 
Fig. 1.

\begin{figure*}[h]
\resizebox{\hsize}{!}{\includegraphics{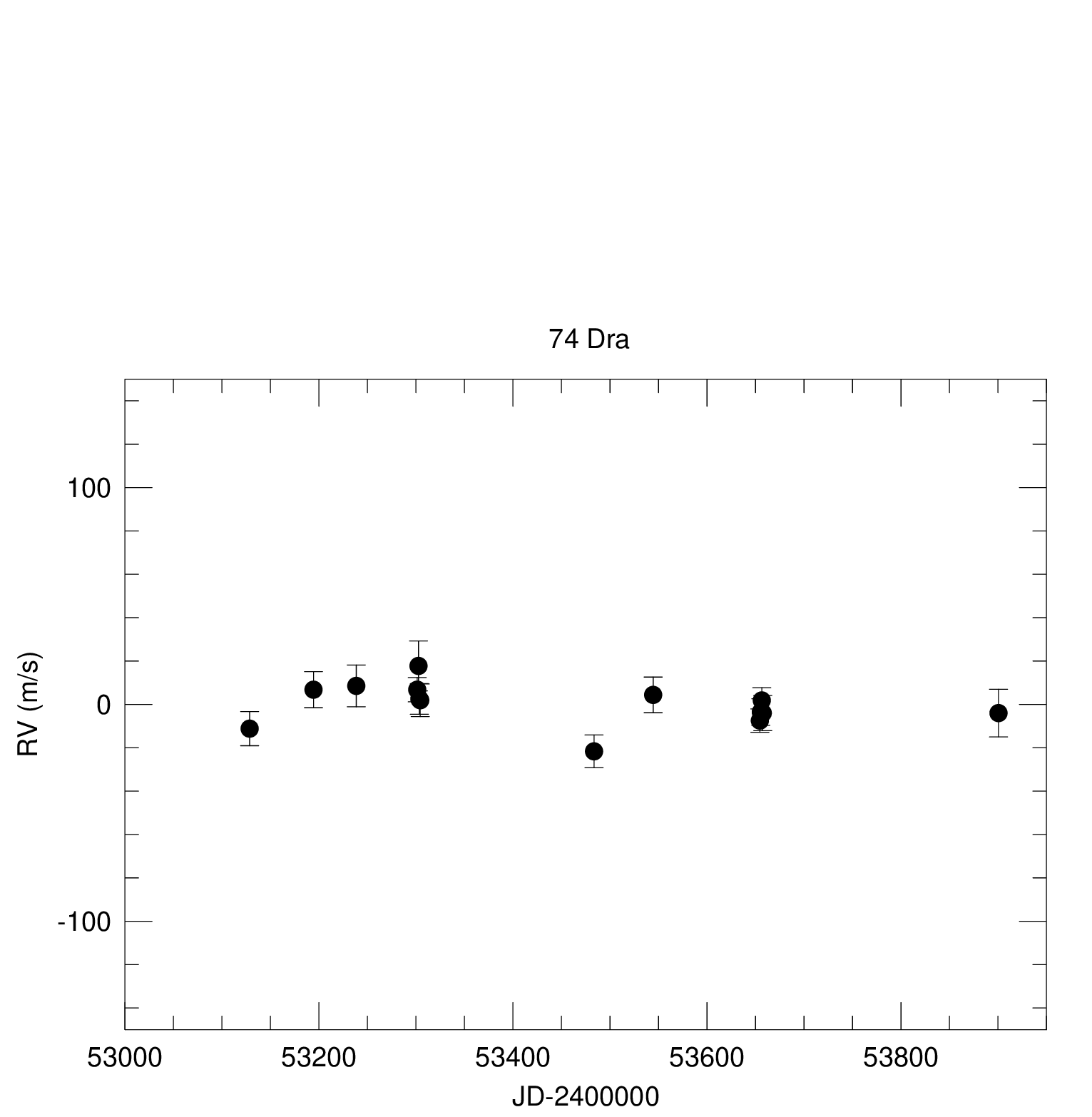}}
\caption{RV measurements of the target star 74 Dra. Given a strict RV limit of $<$ 10 m\,s$^{-1}$ this star is per definition ``constant'' and used as RV standard.} 
\end{figure*}

The time analysis of the measured  H$\alpha$ activity will be
discussed in more detail for both $TLS$ planet candidates in \S3.1 and \S3.2.


\section{Properties of the stars 42~Dra and HD~139357 and of their planetary 
companions}
\subsection{42 Dra}
42 Dra (= HD 170693 =  HR 6945 = HIP 90344) has a visual magnitude of 
$V$~=~4.83~mag and is classified in $SIMBAD$ as K1.5~III star. As for the 
$HIPPARCOS$ parallaxes, we adopt the values derived by van Leeuwen (2007).
The corresponding value is 10.36~$\pm$~0.20~mas and this implies an 
absolute magnitude M$_{\mathrm{V}}$~=~$-$0.09~$\pm$~0.04~mag.
As demonstrated by van Leeuwen $\&$ Fantino (2005) and van Leeuwen (2007), this 
new reduction has
substantially improved the results for the parallaxes of bright stars, with
respect to the original values in the ESA catalogue (1997). In the specific
case of 42~Dra and HD 139357, although the mean parallax values are close to
their previous values, their standard errors are now reduced by factors of
about 2.\\ 
The stellar parameters of 42 Dra, summarized in Tab. 1 and shown in Fig. 2, 
were either obtained 
from the literature or derived from our analysis of the stellar spectra taken 
without the iodine cell. These high-quality templates allowed us to determine 
accurate Fe abundances [Fe/H], effective temperatures T$_{\mathrm{eff}}$, 
logarithmic surface gravities $\log g$, and microturbulence velocities $\xi$.\\
The results of this analysis for 42~Dra, HD 139357, and the rest
of the Tautenburg sample as well as a comparison with previous studies will be 
presented in more detail in a forthcoming paper.

\begin{table}[h]
\caption{Stellar parameters of 42 Dra. The properties of the host star are listed in detail.}
\vspace{-0.5cm}
$$
\begin{array}{lll}
\hline
\hline
\mathrm{Spectral\,\,type}       & \mathrm{K1.5III}        & $HIPPARCOS$\\
m_{\mathrm{V}}            & 4.833 \pm 0.005      & [\mathrm{mag}] \\
M_{\mathrm{V}}            & -0.09 \pm 0.04     & [\mathrm{mag}] \\
B-V                     & 1.19 \pm 0.005       & [\mathrm{mag}] \\
\mathrm{Parallax}                    & 10.36 \pm 0.20       & [\mathrm{mas}] \\
\mathrm{Distance}                    & 96.5 \pm 1.9       & [\mathrm{pc}] \\
M_{\mathrm{*}}^{(a)}             & 0.98 \pm 0.05     & [\mathrm{M_{\odot}}] \\
R_{\mathrm{*}}^{(a)}             & 22.03 \pm 1.00       & [\mathrm{R_{\odot}}] \\
\mathrm{Age}^{(a)}               & 9.49 \pm 1.76      & [\mathrm{Gyr}] \\
T_{\mathrm{eff}}^{(a)}  & 4200 \pm 70          & [\mathrm{K}] \\
\mathrm{[Fe/H]}^{(a)}            & -0.46 \pm 0.05      & [\mathrm{dex}] \\
\log{g}^{(a)}            & 1.71 \pm 0.05         & [\mathrm{dex}] \\
\mathrm{micro\,turbulence}^{(a)}     & 1.6 \pm 0.8             & [\mathrm{{km\,s}^{-1}}]\\
\hline
\hline
\end{array}
$$
{\footnotesize
$^{(a)}$ D\"{o}llinger (2008), D\"{o}llinger (2009), in preparation}
\end{table}

\begin{figure*}[h]
\resizebox{\hsize}{!}{\includegraphics{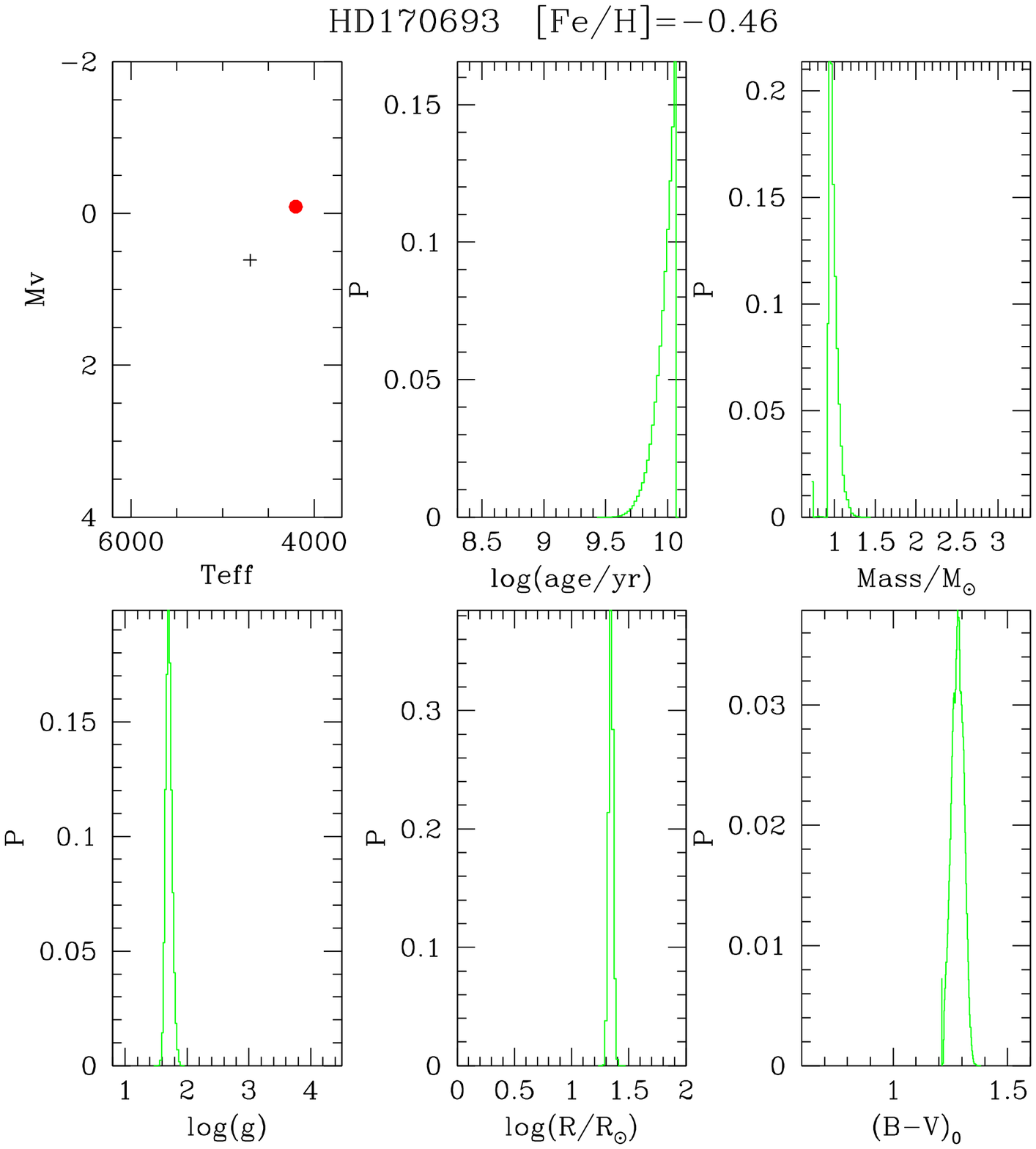}}
\caption{Diagram of the total Probability Distribution Function (PDF) of the stellar parameters for 42 Dra. The peak of the PDF represents the most probable value of each stellar property.}
\end{figure*}

Metallicity, T$_{\mathrm{eff}}$ and absolute $V$-band magnitude, as derived 
from $HIPPARCOS$ parallaxes, were used as input values to estimate 
the mass, age, and radius of each programme star by comparing these values
to theoretical isochrones using a modified version of 
Jo$\!\!\!/$rgensen $\&$ Lindegren's (2005) method. The result of these method
is the total Probability Distribution Function (PDF) for each stellar property
(see Fig. 2). The peak of the PDF represents the most probable value
of a stellar parameter. A detailed description of 
the procedure is given in da Silva et al. (2006).\footnote{The present
implementation of the PDF method is publicly available via the web
interface\\ {\tt http://stev.oapd.inaf.it/param}.}\\

\begin{table}[h]
\caption{Radial velocity measurements for 42 Dra.}
\vspace{-0.5cm}
$$
\begin{array}{lrr}
\hline
\hline
\mathrm{JD}   & \mathrm{RV} [\mathrm{{m\,s}^{-1}}]  & \sigma [\mathrm{{m\,s}^{-1}}]\\
2453128.419501 &     -53.2293   & 3.90\\
2453192.524517  &   -114.8587   & 4.12\\
2453193.403201   &  -126.0224   & 3.31\\
2453238.410453    &  -61.4688   & 2.84\\
2453301.329173     &  50.4884   & 3.96\\
2453302.522409  &     67.6217   & 5.92\\
2453303.416619   &    49.0516   & 4.46\\
2453304.252831    &   59.8095   & 3.59\\
2453304.257380     &  66.2191   & 6.60\\
2453419.712262   &    79.7443   & 3.23\\
2453460.616950    &   98.9271   & 3.98\\
2453461.636258     & 100.0269   & 3.48\\
2453476.524540      & 44.9801   & 2.88\\
2453481.567590       &66.8865   & 4.16\\
2453482.574098     &  80.2309   & 4.36\\
2453483.509160     &  86.7844   & 3.04\\
2453484.527265     &  29.8126   & 3.88\\
2453637.327828     & -88.0594   & 3.52\\
2453654.447141     & -40.0147   & 4.13\\
2453655.432914     & -59.9776   & 3.36\\
2453655.606802     & -41.1324   & 4.49\\
2453656.461186     & -74.5463   & 3.53\\
2453656.620815     & -65.0586   & 3.79\\
2453657.433509     & -72.2340   & 3.77\\
2453657.609979     & -66.5591   & 4.84\\
2453899.530210     &  64.2880   & 5.17\\
2453900.481314     &  88.1803   & 5.82\\
2453901.570729     &  91.9168   & 4.53\\
2453904.502606     & 116.6142   & 5.73\\
2453905.551268     & 129.6684   & 5.48\\
2453840.546573     &  88.7195   & 3.29\\
2453995.339307     &  81.0351   & 3.00\\
2454041.290860     &  -5.0989   & 3.74\\
2454047.266968     &   4.5754   & 6.48\\
2454097.222261     & -67.5189   & 4.13\\
2454136.697758     &-139.7974   & 4.20\\
2454157.661375     &-125.1538   & 3.45\\
2454099.192333    &  -63.6536   & 4.43\\
2454099.196558    &  -66.4595   & 5.35\\
2454192.644336    & -113.0858   & 3.29\\
2454171.672927    & -193.1661   & 3.40\\
2454309.380979    &   49.9884   & 3.23\\
2454313.374845    &   86.8409   & 3.47\\
2454330.329069    &  101.4740   & 3.60\\
2454337.334089    &   86.0530   & 3.15\\
\hline
\hline
\end{array}
$$
\end{table}

\begin{figure*}[h]
\resizebox{\hsize}{!}{\includegraphics{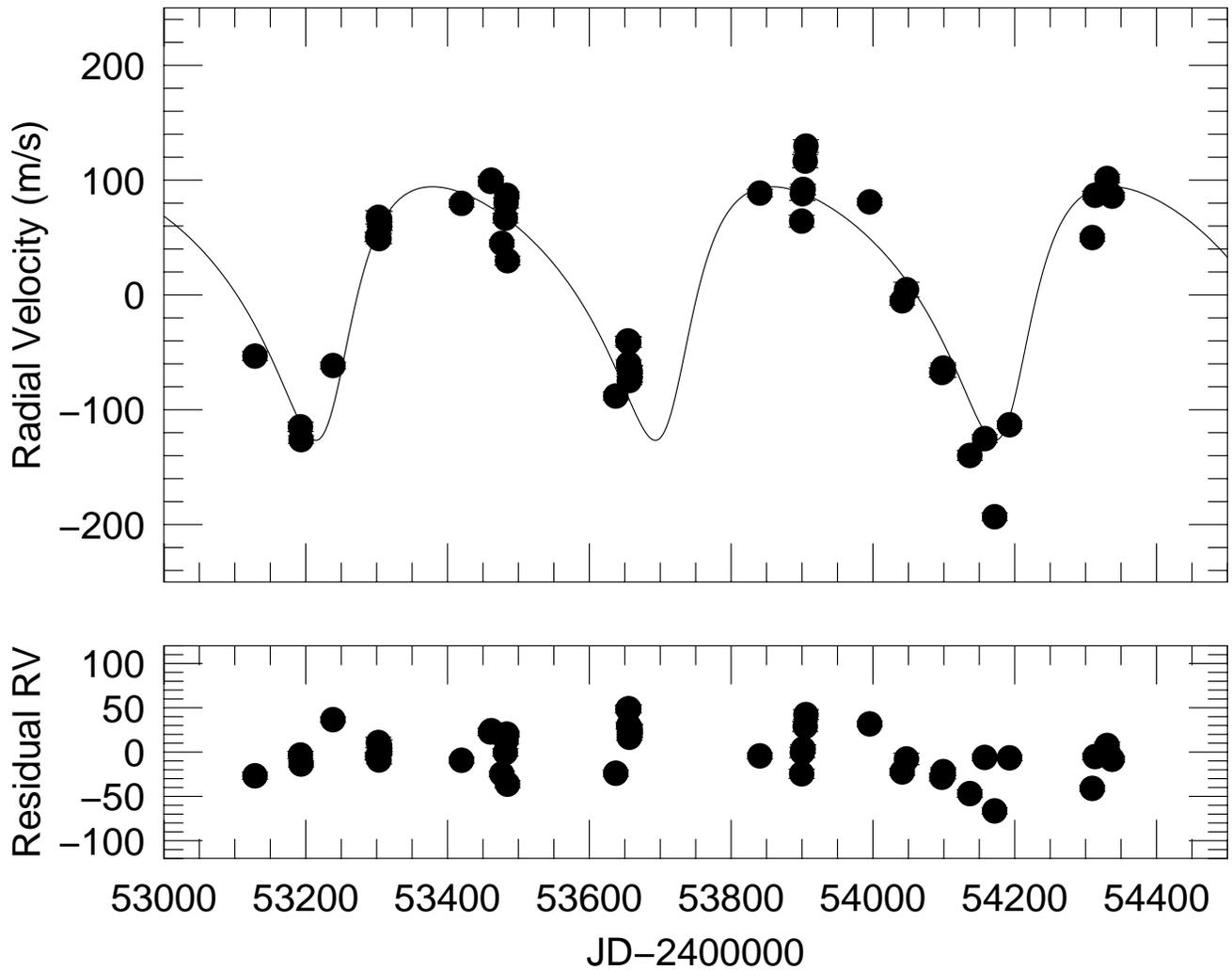}}
\caption{Radial velocity measurements for 42 Dra (top). The solid line is the orbital solution. The RV residuals (lower) seem to contain short-period variations. However the Lomb-Scargle periodogram shows no significant frequency.}
\end{figure*}

\begin{figure*}[h]
\resizebox{\hsize}{!}{\includegraphics{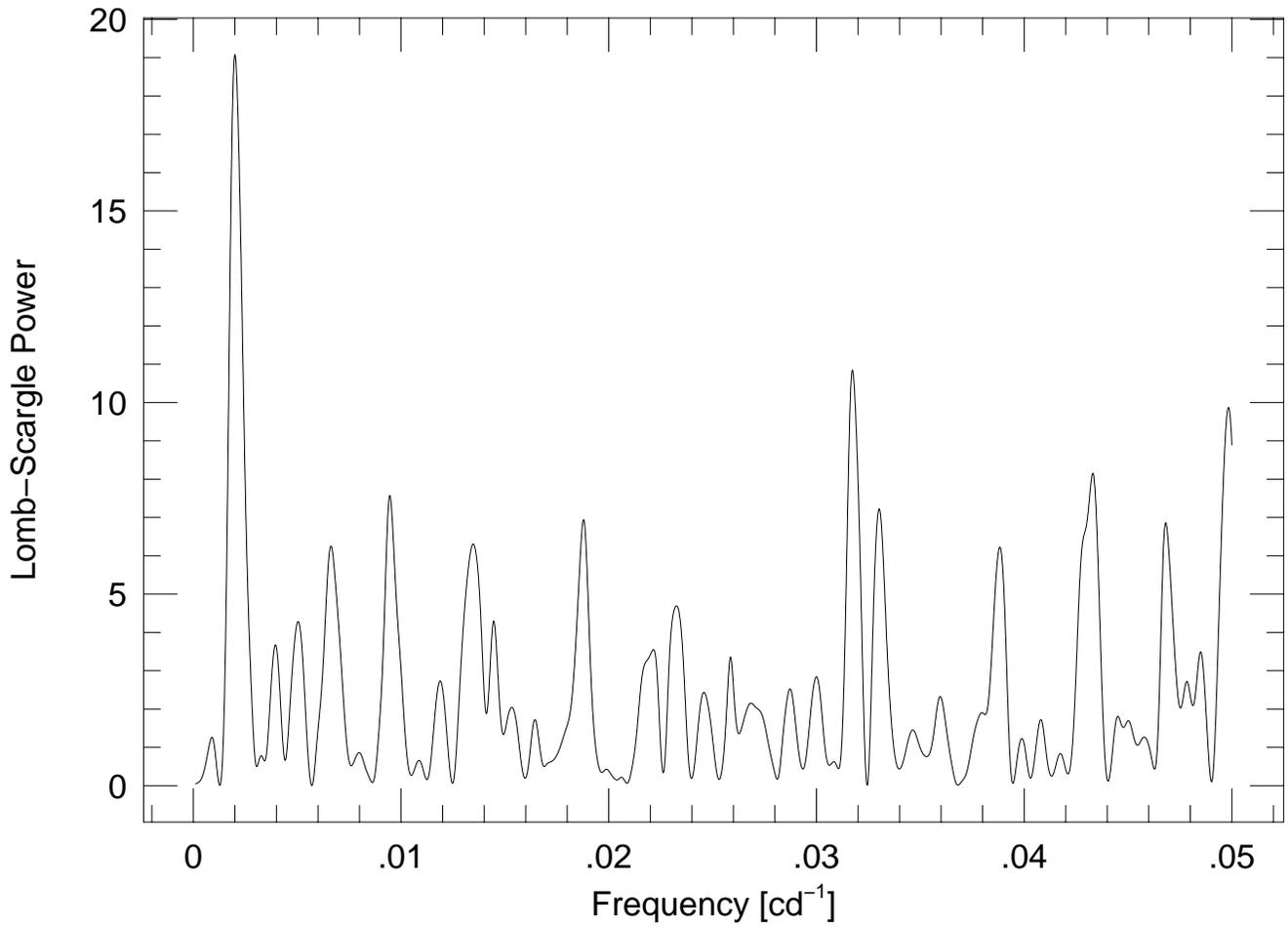}}
\caption{Lomb-Scargle periodogram for 42~Dra. There is a very high peak with the scargle power 19.0 at a frequency $\nu$~=~0.00196~c\,d$^{-1}$ corresponding to a period of 510.2~days.
}
\end{figure*}

\begin{figure*}[h]
\resizebox{\hsize}{!}{\includegraphics{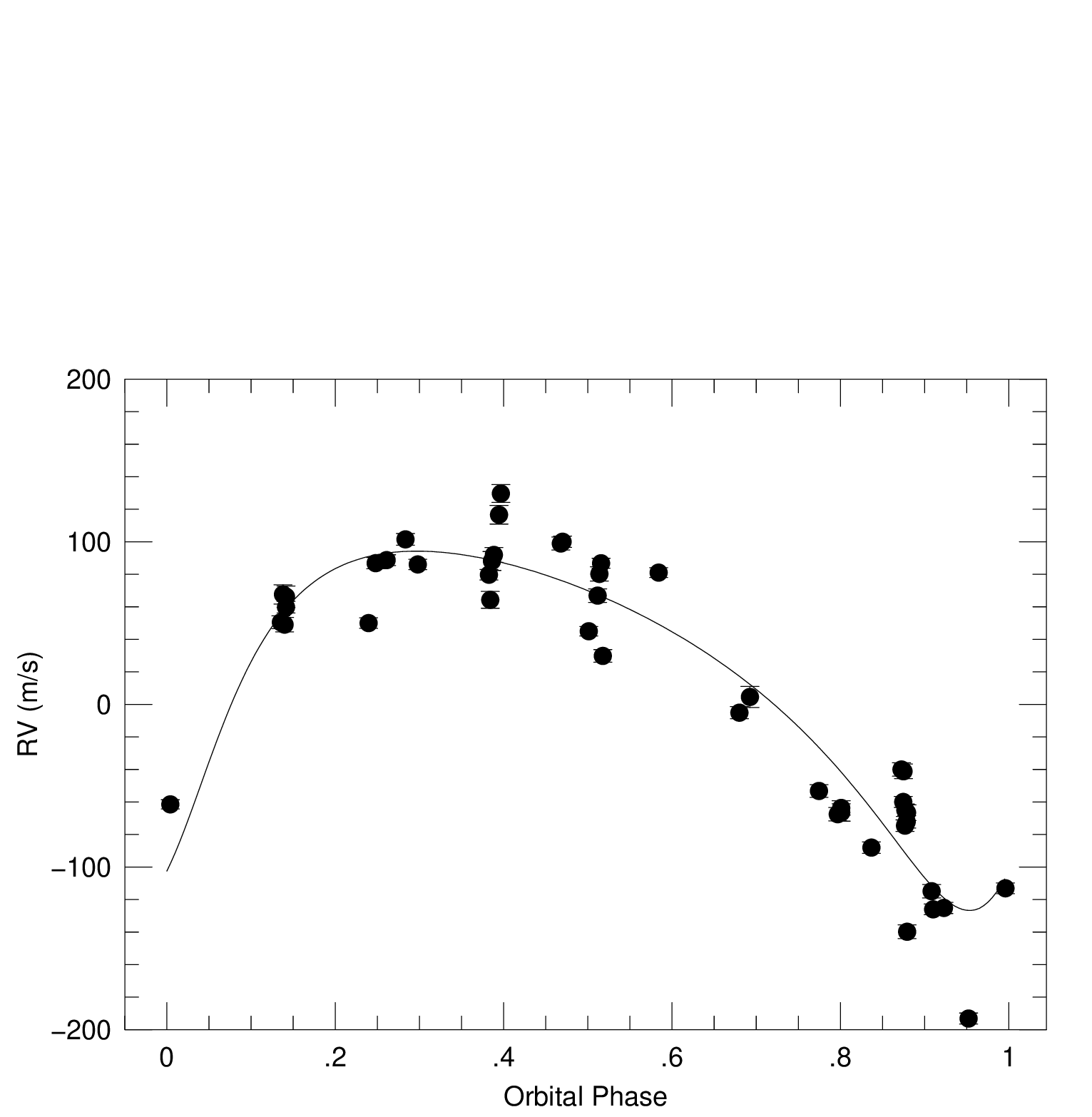}}
\caption{Radial velocity measurements for 42 Dra phased to the orbital period. 
}
\end{figure*}

The time series of our RV measurements (see Tab. 2) of 42 Dra 
including the orbital fit is plotted in the upper part of Fig. 3. The RV curve 
of the K giant 42~Dra shows an obvious sinusoidal variation. With the exception of one point (lower right), possibly caused by stellar oscillations, the 
orbital solution fits the data well. 
The  Lomb-Scargle periodogram of the RV measurements is shown in Fig. 4 and
this confirms the presence of a statistically significant 
peak at a frequency of $\nu$~=~0.00196~c\,d$^{-1}$ ($P$ = 510.2 days). 

The ``False Alarm Probability'' (FAP) of the 510-day period peak using the 
prescription in Scargle (1982) is estimated to be 
$\approx$~2~$\times$~10$^{-7}$.  The FAP was also estimated using a
``bootstrap randomization technique'' (see K\"urster et al. 1999).
The RV values were randomly shuffled
keeping the times fixed and a Lomb-Scargle periodogram calculated for
each random data set over the same frequency range shown in Fig. 4.
After 200{,}000 ``shuffles'' there was no instance of the random periodogram
having power larger than the real data set. This confirms that the periodic
signal is not due to noise or data sampling.

An orbital solution to the data was  made using the general non-linear
least squares programme $GaussFit$ (McArthur et al. 1994). This resulted
in a shorter period of 479.1~$\pm$~6.2~days.
The discrepancy between the periods is due to the fact that the Lomb-Scargle 
programme uses sine functions (i.e. a circular orbit) to find periodic
signals in the data. The orbital eccentricity of 42 Dra b
is $e$~=~0.38~$\pm$~0.06, so a pure sine wave is not the best fit
to the data. Consequently, the Lomb-Scargle periodogram should yield
a less reliable period.

All orbital elements are listed in Tab.~3. The corresponding mass function is
$f(m)$~=~(5.29~$\pm$~0.62)~$\times$~10$^{-8}$~M$_{\odot}$.
Using our derived stellar mass of 0.98~$\pm$~0.05~M$_{\odot}$
we calculated a ``minimum mass'' of $m~\sin~i$~=~3.88~$\pm$~0.85~M$_{\mathrm{Jup}}$
for the companion.
The line in Fig.~3 shows the orbital solution to the RVs. Fig.~5 shows the 
phase-folded RV variations and orbital fit.

\begin{table}[h]
\caption{Summary of all orbital parameters for the planetary companion to 42 Dra.}
\vspace{-0.5cm}
$$
\begin{array}{lll}
\hline
\hline
\mathrm{Period} [\mathrm{days}]        & 479.1 \pm 6.2 \\
T_{\mathrm{periastron}}[\mathrm{JD}]   & 2452757.4 \pm 3.7\\
K [\mathrm{{m\,s}^{-1}}]                 & 110.5 \pm 7.0\\
\sigma(\mathrm{O-C)}) [\mathrm{{m\,s}^{-1}}]       & 26.0 \\
e                                      & 0.38 \pm 0.06\\
\omega [\mathrm{deg}]                           & 218.7 \pm 10.6\\
f(m) [\mathrm{M_{\odot}}]                       & (5.29 \pm 0.62) \times 10^{-8}\\
a [\mathrm{AU}]                        & 1.19 \pm 0.01\\
\hline
\hline
\end{array}
$$
\end{table}

The RV residuals of 42 Dra are plotted in the lower panel
of Fig. 3. A periodogram 
analysis of the residual RV data showed no
additional significant frequencies in the data.
To check, if rotational modulation might
be the cause of the observed RV variations we analyzed the $HIPPARCOS$
photometry and the H$\alpha$ activity of 42~Dra.
Figs. 6 and 7 show the periodogram of the photometry and
the H$\alpha$ activity, respectively. Neither the $HIPPARCOS$ photometry
nor the H$\alpha$ data show significant variations at the orbital frequency.
From the projected rotational velocity $v~\sin~i$~$<$~1.0~km\,s$^{-1}$
(de Medeiros et al. 1996) and the adopted stellar radius, listed in Tab. 1, 
we have estimated the lower limit of the rotational period. This calculated
value of 1113 days is completely different from the orbital period
of the planetary companion (see Tab. 3) and confirms that the RV variations 
are not due to rotational modulation. The typical lifetime of active zones 
for K giants in general can range between hours to around several hundreds 
of days.

\begin{figure*}[h]
\resizebox{\hsize}{!}{\includegraphics{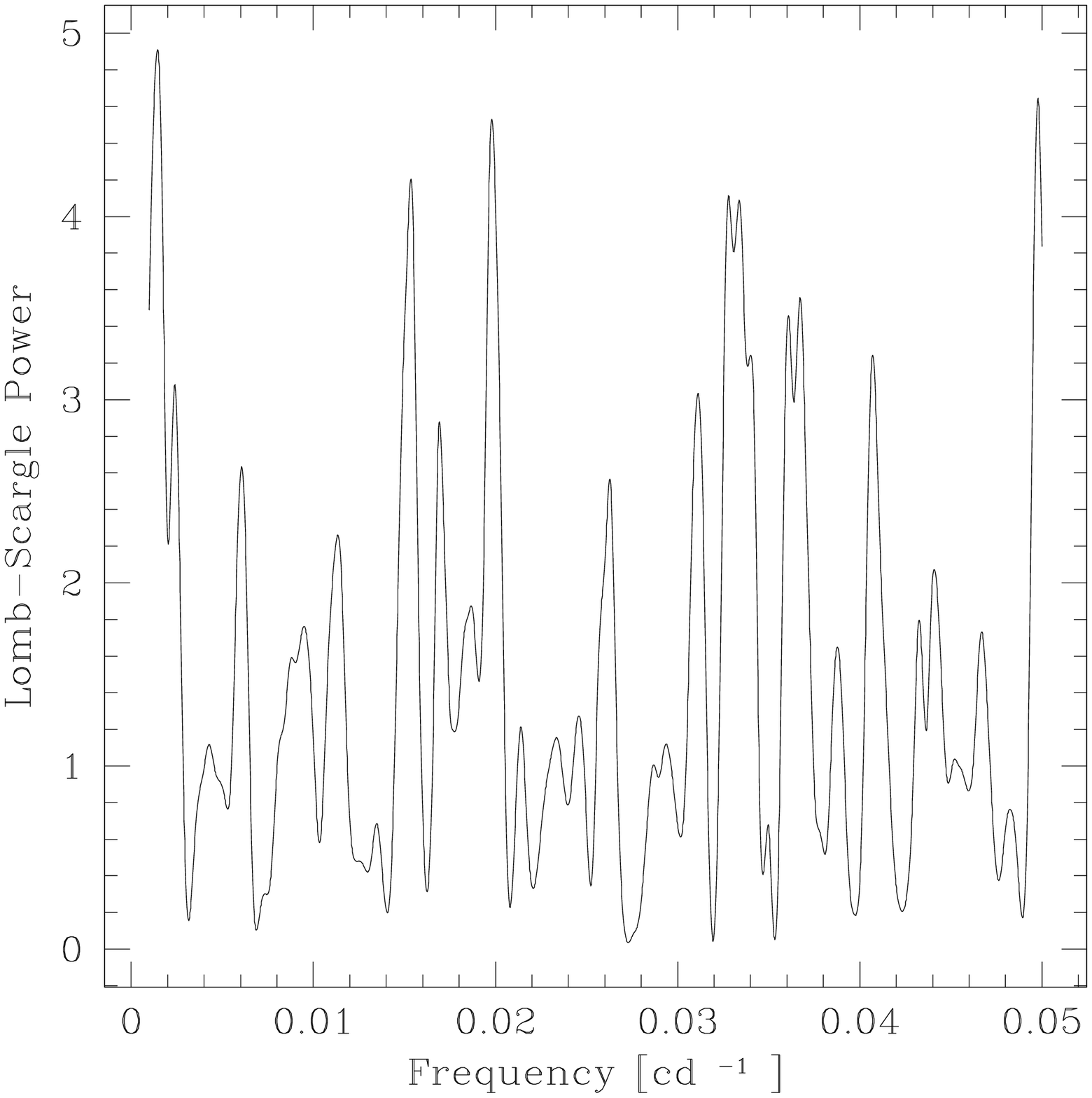}}
\caption{Lomb-Scargle periodogram of the $HIPPARCOS$ photometry for 42 Dra.
}
\end{figure*}

\begin{figure*}[h]
\resizebox{\hsize}{!}{\includegraphics{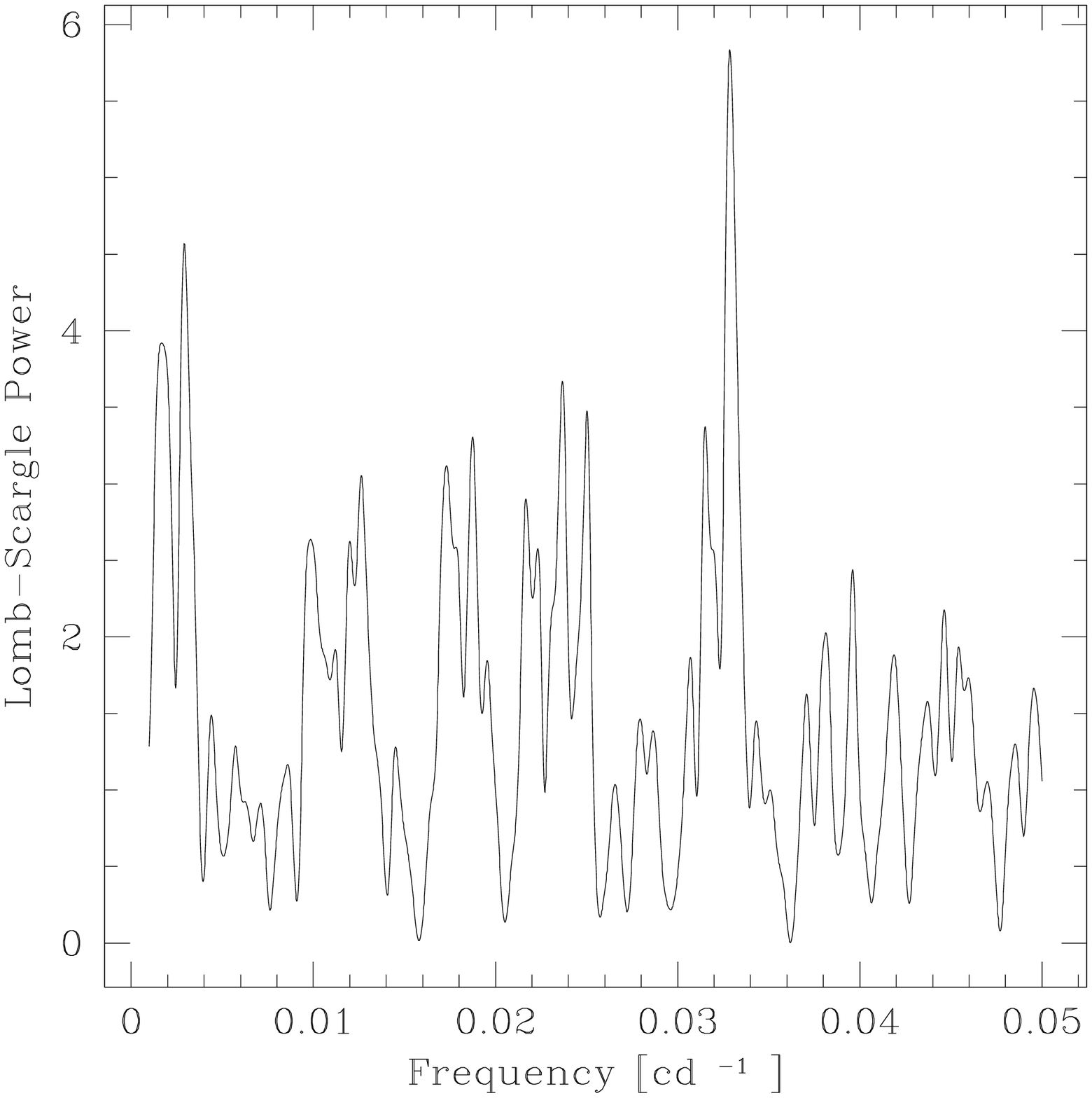}}
\caption{Lomb-Scargle periodogram of the 42 Dra H$\alpha$ variations.
}
\end{figure*}

We note that the sigma of the RVs around the orbital solution is about
26~m\,s$^{-1}$, significantly higher than our RV precision. There
is also one point which differs significantly from the predicted RV
of the orbital solution (lower right in Fig.~3, near the third RV minimum).
This scatter most likely arises from stellar oscillations. It is well known
that K giant stars 
show RV variations due to p-mode oscillations with periods ranging from hours 
to days (e.g. Hatzes $\&$ Cochran 1995; de Ridder et al. 2006; Hatzes $\&$
Zechmeister 2007). We can use the scaling relations of 
Kjeldsen $\&$ Bedding (1995) to estimate 
the expected amplitude of such variations. The velocity amplitude 
for stellar oscillations is predicted to be v$_{osc}$~=~((L/L$_{\odot}$)/(M/M$_{\odot}$))$\times$23.4~cm\,s$^{-1}$ according to Kjeldsen $\&$ Bedding (1995) Eq.~7.
Using the luminosity, calculated with the spectroscopic effective temperature
T$_{\mathrm{eff}}$, the stellar radius R$_{*}$, and the stellar mass M$_{*}$
for 42 Dra results in an amplitude of
22.9~m\,s$^{-1}$, consistent with our rms scatter. The characteristic periods
for such oscillations are expected to be about 1.7~days according to the
Kjeldsen $\&$ Bedding scaling relations. Thus, the scatter of the RV data
about the orbital solution can be accounted for by stellar oscillations.

\subsection{HD 139357}
Another exoplanet orbits the slightly metal-poor giant HD~139357
(=  HR~5811 = HIP~76311) having the longest period of all Tautenburg planet 
candidates within this K giant
star sample with a period comparable with the duration of our survey 
The stellar parameters of this star are summarized in Tab.~4. The PDF
diagrams of the stellar parameters are shown in Fig.~8.

\begin{figure*}[h]
\resizebox{\hsize}{!}{\includegraphics{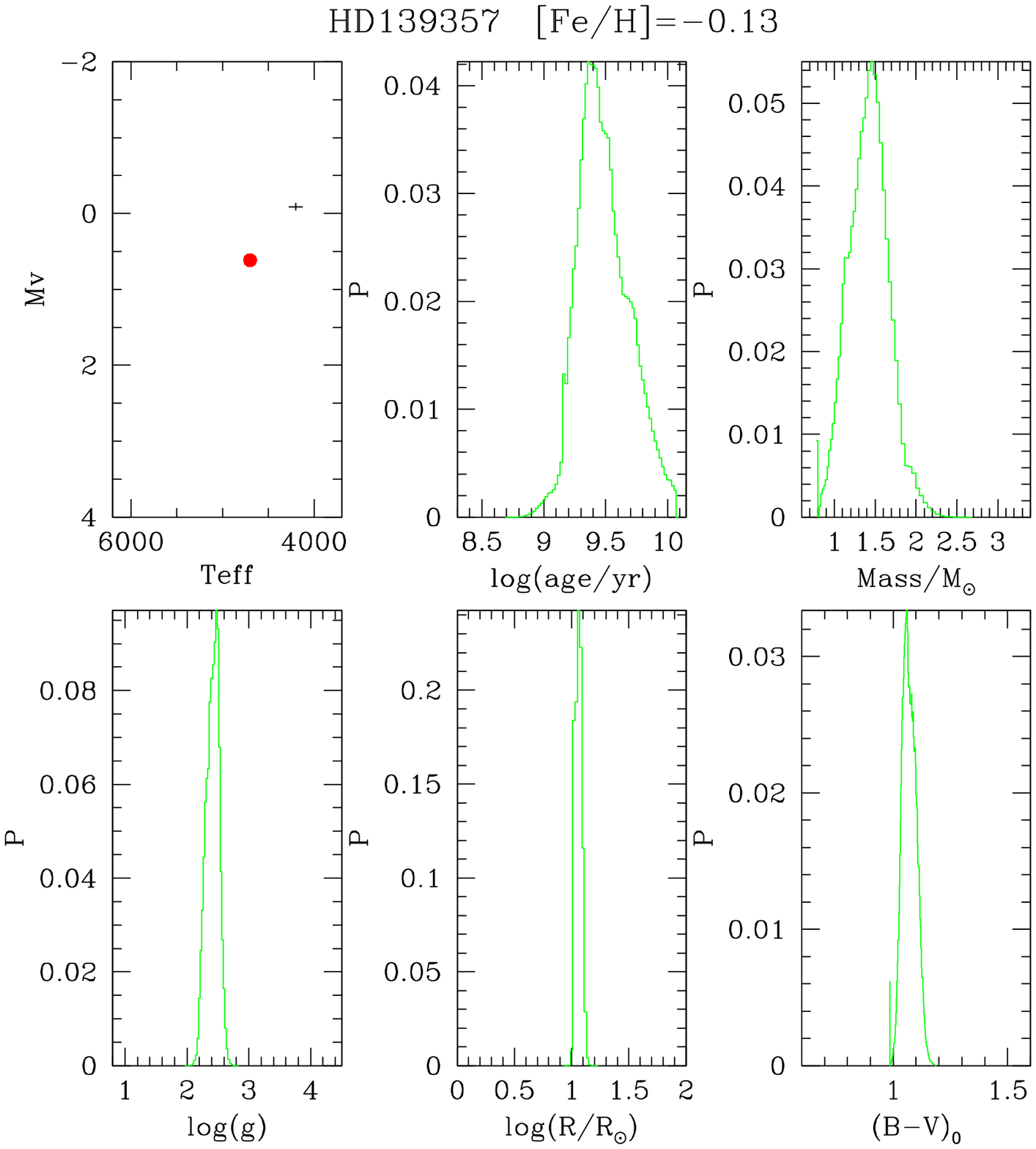}}
\caption{Stellar parameters of HD~139357. The properties of the host star are listed in detail.}

\end{figure*}

The time series of the
RV measurements for HD 139357 is shown in the upper panel
of Fig.~9. The lower panel shows the RV residuals after subtracting the
orbital solution (see below).

\begin{table}[h]
\caption{The values of the stellar properties of the planet host star HD 139357 are listed.}
\vspace{-0.5cm}
$$
\begin{array}{lll}
\hline
\hline
\mathrm{Spectral\,\,type}       & \mathrm{K4III}        & $HIPPARCOS$\\
m_{\mathrm{V}}       & 5.977 \pm 0.005        & [\mathrm{mag}] \\
M_{\mathrm{V}}       & 0.61 \pm 0.08        & [\mathrm{mag}] \\
B-V                 & 1.2 \pm 0.005        & [\mathrm{mag}] \\
\mathrm{Parallax}               & 8.46 \pm 0.30          & [\mathrm{mas}] \\
\mathrm{Distance}      & 118.2 \pm 4.4         & [\mathrm{pc}]  \\
M_{\mathrm{*}}^{(a)}                 & 1.35 \pm 0.24   & [\mathrm{M_{\odot}}] \\
R_{\mathrm{*}}^{(a)}                & 11.47 \pm 0.75   & [\mathrm{R_{\odot}}] \\
\mathrm{Age}^{(a)}                    & 3.07 \pm 1.47        & [\mathrm{Gyr}] \\
T_{\mathrm{eff}}^{(a)}     & 4700 \pm 70            & [\mathrm{K}] \\
\mathrm{[Fe/H]}^{(a)}         & -0.13 \pm 0.05        & [\mathrm{dex}] \\
\log{g}^{(a)}          & 2.9 \pm 0.15           & [\mathrm{dex}] \\
\mathrm{micro\,turbulence}^{(a)}      & 1.6 \pm 0.8            & [\mathrm{{km\,s}^{-1}}]\\
\hline
\hline
\end{array}
$$
{\footnotesize
$^{(a)}$ D\"{o}llinger (2008), D\"{o}llinger (2009), in preparation}
\end{table}

\begin{table}[h]
\caption{Radial velocity measurements for HD 139357.}
\vspace{-0.5cm}
$$
\begin{array}{lrr}
\hline
\hline
\mathrm{JD}   & \mathrm{RV} [\mathrm{{m\,s}^{-1}}]  & \sigma [\mathrm{{m\,s}^{-1}}] \\
2453051.723680   &      54.6949   &  6.84\\
2453123.549508    &    23.4178   &  6.88\\
2453124.560375     &   25.5474   &  6.88\\
2453128.560125      &  10.4159   &  6.7\\
2453189.548814     &  -10.5455   &  6.26\\
2453193.487180    &   -35.7487   &  7.86\\
2453234.514517    &   -40.2842   &  18.80\\
2453234.519540    &   -67.3711   &  8.47\\
2453238.371353    &   -60.0982   &  6.65\\
2453301.406657   &   -126.0780   &  8.43\\
2453303.224983   &    -90.0338   & 8.63\\
2453460.552279   &   -161.6434   & 7.96\\
2453461.541417   &   -167.5064   & 7.89\\
2453477.430919  &    -158.2766   & 6.81\\
2453477.445942   &   -157.2085   & 5.34\\
2453481.533417   &   -158.9196   & 8.19\\
2453483.475855   &   -161.8257   & 6.28\\
2453484.493369   &   -167.2390   & 7.72\\
2453544.479554   &   -134.8390   & 6.68\\
2453636.302387   &    -52.2132   & 7.41\\
2453654.401421   &    -12.7092   & 8.23\\
2453655.391132   &    -42.2359   & 7.65\\
2453656.423032   &    -31.6170   & 8.77\\
2453657.388311   &    -21.9815   & 8.18\\
2453899.494524   &    152.8739   & 7.31\\
2453900.458972   &    131.9554   & 6.75\\
2453901.517734   &    166.2499   & 7.65\\
2453904.481294   &    162.0390   & 7.47\\
2453905.504617   &    166.7464   & 7.09\\
2453905.543261   &    183.4914   & 7.29\\
2453840.461750   &    137.6694   & 6.71\\
2453840.539181   &    138.8760   & 7.12\\
2453861.471208   &     86.9460   & 9.61\\
2453873.541128   &    150.0565   & 10.28\\
2453863.414579   &    126.4515   & 10.88\\
2453995.390302   &    173.9031   & 15.99\\
2454047.208226   &    125.9478   & 7.16\\
2454099.209659   &    128.9890   & 11.13\\
2454099.213884   &    117.1589   & 11.36\\
2454099.213884   &    117.1589   & 11.36\\
2454192.660473   &     71.3733   & 6.05\\
2454171.653737   &     85.5406   & 6.30\\
2454245.373811   &     19.2284   & 6.51\\
2454309.388842   &     -8.9711   & 8.15\\
2454309.393113   &    -30.9135   & 6.67\\
2454317.509380   &    -16.7829   & 6.31\\
2454330.333603   &    -53.7373   & 7.56\\
2454337.326745   &    -35.3002   & 6.24\\
2454338.350758   &    -60.2710   & 7.29\\
\hline
\hline
\end{array}
$$
\end{table}

\begin{figure*}[h]
\resizebox{\hsize}{!}{\includegraphics{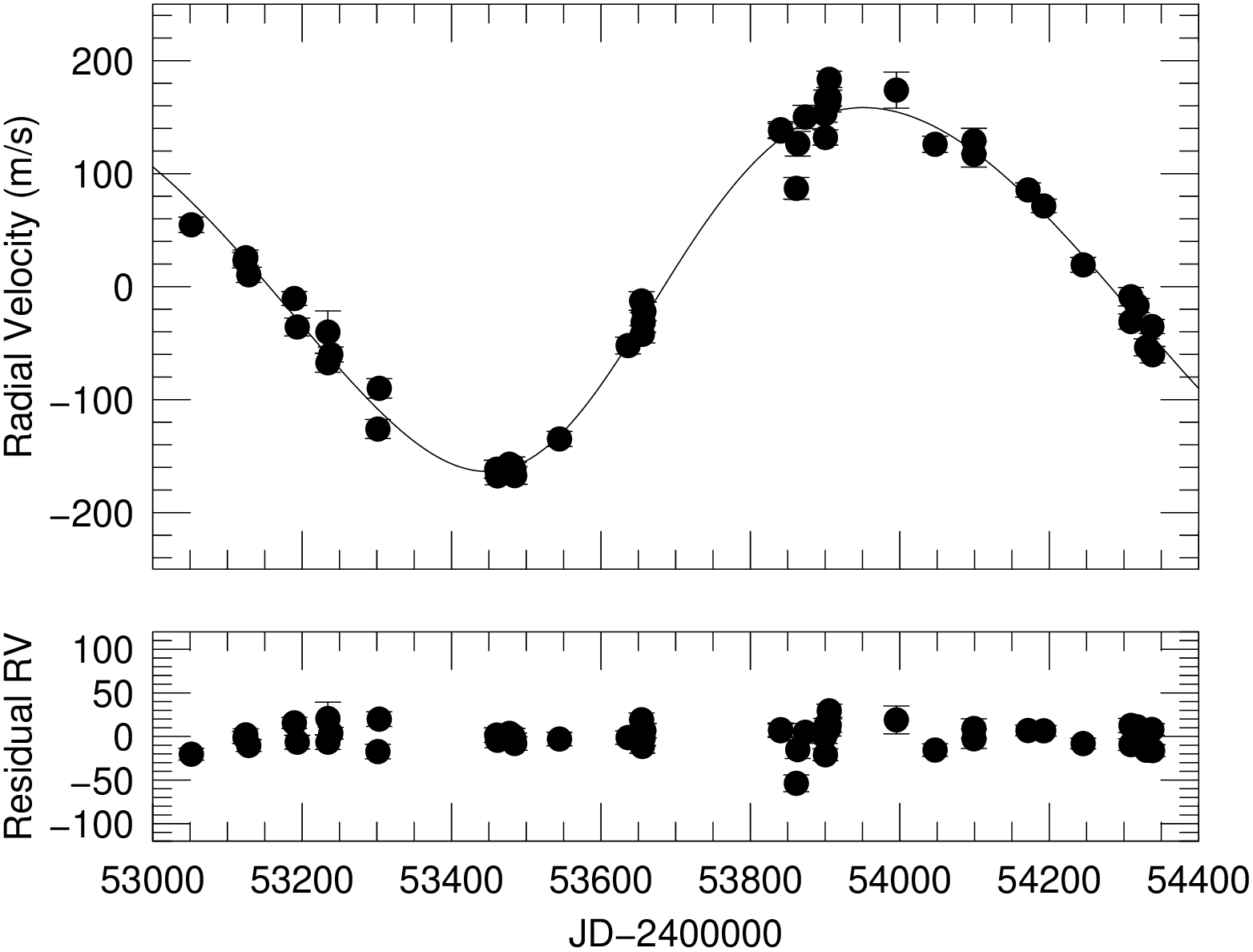}}
\caption{Radial velocity measurements and RV residuals for HD~139357.
}
\end{figure*}

The radial velocity data reveal a nearly sine-like long-period variation. A 
Lomb-Scargle periodogram of the RV data (see Tab. 5) shows significant power 
corresponding to a period of 1124 days.
The FAP of this peak is $\approx$~10$^{-9}$. We do not show the
Lomb-Scargle periodogram as it is readily apparent from Fig.~9 that there
are periodic RV variations in this star.

An orbital solution yielded a revised period of 1125.7 days and a slightly 
eccentric orbit, $e$~=~0.10~$\pm$~0.02. All the orbital elements are listed in 
Tab.~6 and the orbital solution is shown as a line in Fig. 9. The mass function
(see Tab.~6) delivers a ``minimum mass'' of
$m~\sin~i$~=~9.76~$\pm$~2.15~M$_{\mathrm{Jup}}$ for the substellar companion.
It appears that our data window covered slightly more than one orbital period.

\begin{table}[h]
\caption{All orbital parameters for the companion to the K giant host star HD 139357.}
\vspace{-0.5cm}
$$
\begin{array}{lll}
\hline
\hline
\mathrm{Period} [\mathrm{days}]          & 1125.7 \pm 9.0\\
T_{\mathrm{periastron}} [\mathrm{JD}]             & 2452466.7 \pm 3.2\\
K [\mathrm{m\,s^{-1}}]                     & 161.2 \pm 3.2\\
\sigma(\mathrm{O-C)}) [\mathrm{m\,s^{-1}}]        & 14.14\\
e                                 & 0.10 \pm 0.02 \\
\omega [\mathrm{deg}]                    & 235.4 \pm 10.6 \\
f(m) [\mathrm{M_{\odot}}] & (4.79 \pm 0.57) \times 10^{-7} \\
a [\mathrm{AU}]                            & 2.36 \pm 0.02\\
\hline
\hline
\end{array}
$$
\end{table}

The orbital fit to the data is again quite good with an 
rms scatter of only 14~m\,s$^{-1}$. 
The Kjeldsen $\&$ Bedding relations predict an
RV amplitude for stellar oscillations of 13.7~m\,s$^{-1}$ for this star, 
consistent with the rms scatter about the orbit. The Lomb-Scargle periodogram
after removal of the orbital contribution due to the
exoplanet shows no significant periodic signals.


To test if the RV period could arise from rotational 
modulation we searched for significant
frequencies in the 106 data points of the $HIPPARCOS$ photometry.
Fig.~10
shows the Lomb-Scargle periodogram of the photometry.
There is no significant power at the RV frequency.

\begin{figure*}[h]
\resizebox{\hsize}{!}{\includegraphics{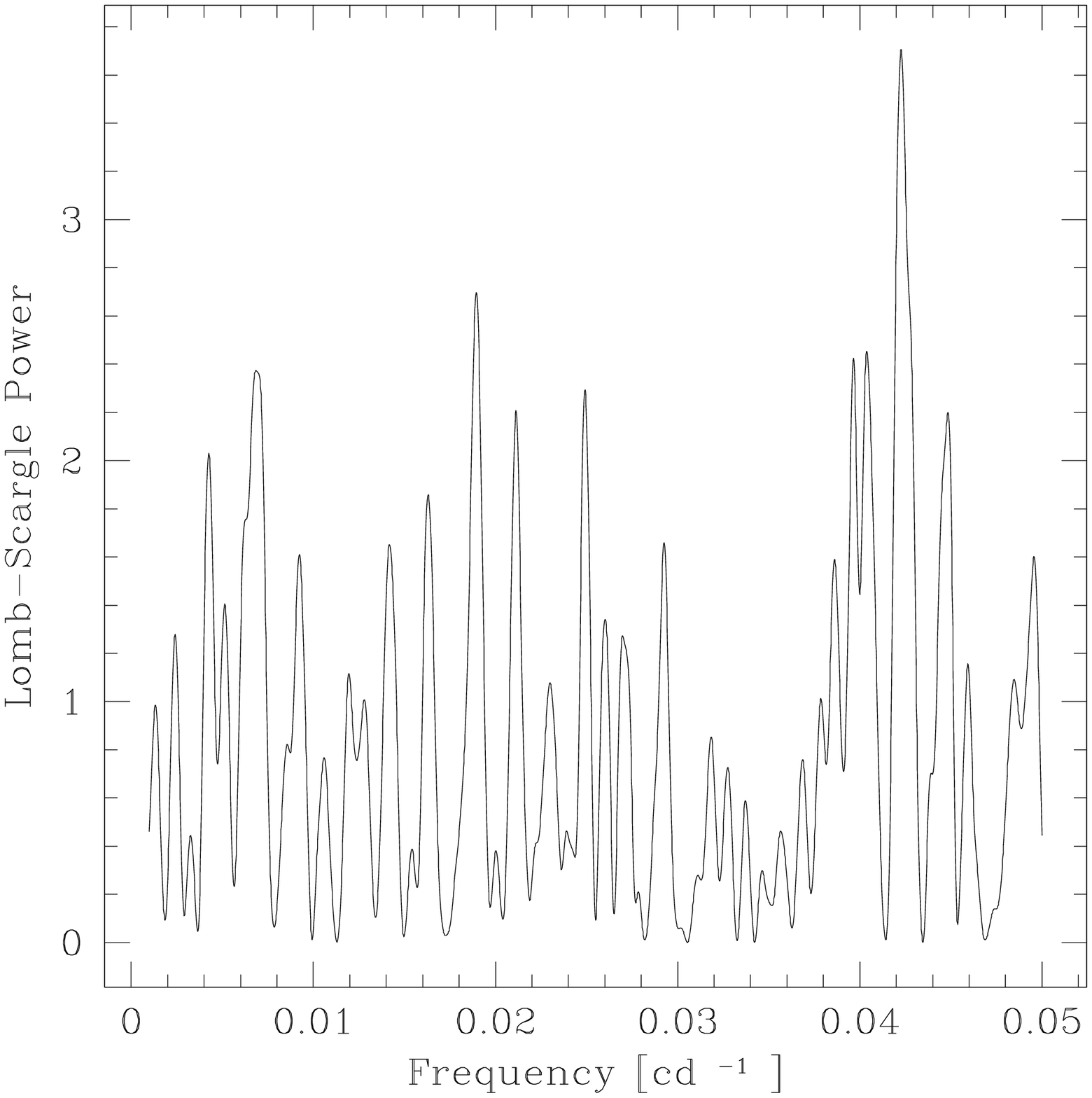}}
\caption{Lomb-Scargle periodogram of the $HIPPARCOS$ photometry for
 HD 139357.
}
\end{figure*}

\begin{figure*}[h]
\resizebox{\hsize}{!}{\includegraphics{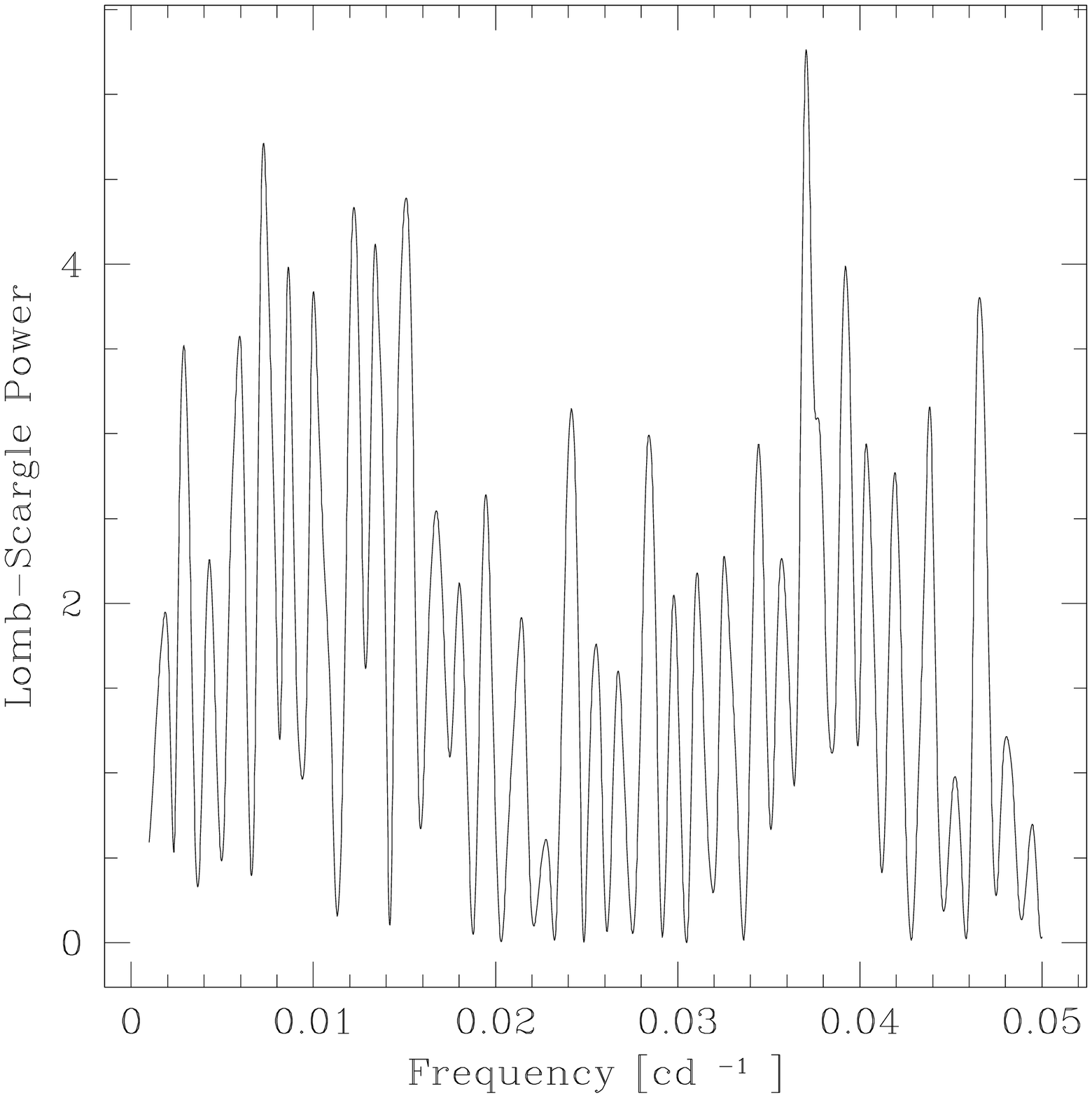}}
\caption{Lomb-Scargle periodogram of the H$\alpha$ variations for HD 139357.
}
\end{figure*}

Fig.~11 shows the Lomb-Scargle periodogram
of the H$\alpha$ variations. Again, there is no significant power near the
RV frequency.

\section{Discussion}
Our radial velocity measurements indicate that the K giant stars 42 Dra and 
HD 139357 host extrasolar giant planets.
In the case of 42~Dra, the planet orbit is modestly eccentric which may argue 
in favour of Keplerian motion. However, it is not highly eccentric
as it is the case for $\iota$~Dra~b. It is still possible for surface structure
to produce such a RV curve. The eccentricity argument for HD~139357 is
less effective due to its nearly circular orbit.
Nevertheless, the lack of 
variability in the $HIPPARCOS$ and the H$\alpha$ data for both
stars is more consistent with the planet hypothesis.
We caution the reader that the $HIPPARCOS$ photometry was 
not simultaneous with our RV measurements. Thus, we cannot exclude that spots 
were not present when $HIPPARCOS$ observed these stars, 
but are present now and are causing 
the RV variations. Anyhow, our H$\alpha$ measurements were made simultaneously
to the RV data. We therefore believe that 
the detected periods of several hundred days in both stars are not due to 
rotational modulation, but rather to planetary companions.\\
In the search for exoplanets it is important to investigate, if there are 
differences between discovered planet-hosting K giants and MS stars and
how this different behaviour of MS and giant stars is correlated with the 
process of planet formation.\\
The first main difference between MS stars and giants is, that
giants on average have a somewhat higher mass than MS stars surveyed for
exoplanets. Taking into account that the mass determination in giant stars
suffers large uncertainties, the masses of planet-hosting giants vary between
$\sim$~0.9 and $\sim$~3 M$_{\odot}$
(da Silva et al. 2006; D\"ollinger 2008), while those of MS
stars are between $\sim$~0.75 and $\sim$ 1.5~M$_{\odot}$.\\
Another difference between MS and giant stars is their large radius in
comparison to MS stars. To study this point we used the stars of the 
$TLS$ programme and the previous southern
giant survey carried out with the Fiber-fed Extended Range Optical
Spectrograph ($FEROS$) by Setiawan et al. (2004a, 2005). The
$FEROS$ and $TLS$ giants have radii on average about 10 times larger than solar
stars (da Silva et al. 2006; D\"ollinger et al. 2008b).
High metallicity could
favour the migration of the planets towards short-period systems; in this
case, metal-rich stars would have many more short-period planets than
metal-poor stars. These planets would be detected among the MS stars, but no
longer around the giants because the star, enlarging its radius as it
evolves, would
have swallowed up them. It was noted in Pasquini et al. (2007) 
(see their Fig.~1)
that the metallicity
distribution is very different for giants than for dwarfs hosting planets with
comparable long orbital periods, that is, excluding those with short orbits.
This would indicate that the effects of migration, even if present, cannot
explain the observed trend.\\
The third difference has been derived from our high resolution spectra.
The abundance analysis for 42~Dra shows that this star is the
most metal-poor of our Tautenburg sample with [Fe/H]~=~$-$0.46~$\pm$~0.05~dex.
However, it is less metal-poor than HD 13189 (Hatzes et al. 2005) with a value 
of $-$0.58~$\pm$~0.05~dex and also less metal-poor than HD 47536 (Setiawan et
al. 2003a) with a value of $-$0.68~$\pm$~0.05~dex. This is interesting because 
MS stars hosting exoplanets tend to be metal-rich compared to stars that do not
posses exoplanets (Santos et al. 2004).
Other authors (Schuler et al. 2005; da Silva et al. 2006) have also
found evidence that planet-hosting giant stars are metal-poor.
Pasquini et al. (2007) have shown that this characteristic holds for most 
giants hosting planets. We also note that Cochran et al. (2007) found a
giant planet around the MS star HD 155358 with a metallicity of 
[Fe/H]~=~$-$0.68~dex. The basic question is if the increased metallicity of 
planet hosting MS stars enhances planet formation, or whether the high
metallicity is caused by the presence of a planetary system. The
correlation with the metallicity can thus be either cause or effect. The
observed metallicity distribution of both types of host stars of extrasolar
planetary systems should be thus very helpful to choose between the two most
popular planet formation mechanisms: core accretion and
gravitational instability.\\
In the first case, favoured by the core accretion
scenario (Pollack et al. 1996), the stars should be overmetallic down to their
center. Results on MS stars obtained by several groups (Fischer $\&$ Valenti 
2005; Ecuvillon et al. 2006) favour this primordial scenario, where stars are 
born in high metallicity clouds.
In the second case the higher values for the metallicity are not primordial, 
but possibly due to the
pollution of the convective envelope of the star by the infall of planets
and/or planetesimals as well as comets or asteroids (Gonzalez 1998; Laughlin
$\&$ Adams 1997; Gonzalez 2001; Murray et al. 1998; Quillen
$\&$ Holman 2000) onto the star. This pollution can be also caused by the total inward migration of a planet onto the parent star as well as the transfer of
material from the disc to the host star as a result of the migration process
(Goldreich $\&$ Tremaine 1980; Lin et al. 1996) or to the
break-up and infall of one or more planets in multiple systems onto the
surface of the star due to gravitational interactions (Rasio $\&$ Ford 1996)
with other companions. If the star was polluted by the debris of
the planetary system, then only the external layers of the atmosphere of the
star were affected by this pollution (Laughlin $\&$ Adams 1997). Assuming this
scenario, the difference in the mass of the convective envelope
between MS stars and giants would explain why the metal excess
observed in MS stars is not observed among evolved stars. The reason is that
the metal excess produced by this pollution, while visible in the thin
atmospheres of solar-type stars, is completely diluted in the extended, massive
atmospheres of the giants.\\
The hypothesis of pollution in combination with the convective envelope
of evolved stars is a possible explanation for the
different metallicities between MS and giant stars, but other reasons 
like a scenario which includes strong differences in planet formation with 
stellar mass and possibly planet migration is plausible and cannot be
excluded.\\

\section{Conclusions}
We have found hints that the K giant stars 42 Dra and HD 139357 host 
extrasolar planets. Both host stars are metal-poor and seem to indicate, in 
contrast to what is observed among MS stars, that giant planets around
giant stars do not favour metal-rich stars.

\begin{acknowledgements}
We are grateful to the user support group of the Alfred Jensch telescope:
B. Fuhrmann, J. Haupt, Chr. H\"{o}gner, U. Laux, M. Pluto, J. Schiller,
and J. Winkler. We are also grateful to K. Biazzo for the 
IDL\footnote{IDL (Interactive Data Language) 
is a data visualization and analysis platform.} programme extracting the
chromospheric contribution.
This research made use of the SIMBAD database, operated at CDS, Strasbourg,
France. 
\end{acknowledgements}


\begin{thebibliography}{}
\bibitem[bu]{bu} Butler, R.P., Marcy, G.W., Williams, E., McCarthy,
         C., Dosanjh, P., \& Vogt, S.S., 1996, PASP 108, 500
\bibitem[co]{co} Cochran, W.D., Endl, M., Wittenmyer, R.A., Bean, J.L. 2007,
AJ, 665, 1407
\bibitem[da06]{da06} da Silva, L., Girardi, L., Pasquini, L., Setiawan, J., von der L\"{u}he, O., de Medeiros, J.R., Hatzes, A., D\"{o}llinger, M.P., and Weiss, A. 2006, A$\&$A, 458, 603
\bibitem[de06]{de06} de Ridder, J., Barban, C., Carrier, F., Mazumdar, A., Eggenberger, P., Aerts, C., Deruyter, S., Vanautgaerden, J. 2006, A$\&$A, 448, 689  
\bibitem[doe07]{doe07} D\"{o}llinger, M.P., Hatzes, A.P., Pasquini, L., Guenther, E.W., Hartmann, M., Girardi, L., and Esposito, M. 2007, A$\&$A, 472, 649
\bibitem[doe08]{doe08} D\"{o}llinger, M.P. 2008, PhD, LMU M\"{u}nchen, 
accepted
\bibitem[doe09]{doe09} D\"{o}llinger, M.P. 2009, in preparation
\bibitem[ecu]{ecu} Ecuvillon, A., Israelian, G., Santos, N.C., Mayor, M., and Gilli, G. 2006, A$\&$A, 449, 809
\bibitem[esa]{esa} ESA, 1997, yCat.1239....OE
\bibitem[fa]{fa} Fahlman, G.G., and Glaspey, J.W. 1973 in Astronomical Observations with Television Type Sensors, ed. J.W. Glaspey, and G.A.H. Walker, (Vancouver, B.C.: Inst. of Astronomy and Space Science), 347
\bibitem[fi]{fi} Fischer, D., $\&$ Valenti, J. 2005, ApJ, 622, 1102
\bibitem[frei]{frei} Freire Ferrero, R., Frasca, A., Marilli, E., and Catalano, S. 2004, A$\&$A, 413, 657
\bibitem[fr02]{fr02} Frink, S., Mitchell, D.S., Quirrenbach, A., Fischer, D.A., Marcy, G.W., and Butler, R. P. 2002, ApJ, 576, 478

\bibitem[gol]{gol} Goldreich, P., and Tremaine, S. 1980, ApJ, 241, 425
\bibitem[gon98]{gon98} Gonzalez, G. 1998, A$\&$A, 334, 221                      \bibitem[gon01]{gon01} Gonzalez, G., Laws, C., Tyagi, S., and Reddy, B.E. 2001,
AJ, 121, 432                                                                    
\bibitem[hat93]{hat93} Hatzes, A.P., and Cochran, W.D. 1993, ApJ, 413, 339
\bibitem[hat95]{hat95} Hatzes, A.P., and Cochran, W.D. 1995, ApJ, 452, 401
\bibitem[hat05]{hat05} Hatzes, A.P., Guenther, E.W., Endl, M., Cochran, W.D., D\"{o}llinger, M.P., and Bedalov, A. 2005, A$\&$A, 437, 743
\bibitem[ha06]{hat06} Hatzes, A.P., Cochran, W.D., Endl, M., Guenther, E.W., Saar, S.H., Walker, G.A.H., Yang, S., Hartmann, M., Esposito, M., Paulson, D.B., and D\"{o}llinger, M.P. 2006, A$\&$A, 457, 335
\bibitem[ha07]{hat07} Hatzes, A.P., $\&$ Zechmeister, M. 2007, AAS, 211, 2104 
\bibitem[her]{her} Herbig, G.H. 1985, ApJ, 289, 269
\bibitem[jo]{jo} Jo$\!\!\!/$rgensen, and B.R., Lindegren, L. 2005, A$\&$A, 436, 127
\bibitem[johnsona]{johna} Johnson, J.A., Butler, R.P,. Marcy, G.W., Fischer, D.A., Vogt, S.S., Wright, J.T., and Peek, K.M.G. 2007a, ApJ, 670, 833
\bibitem[johnsonb]{johnb} Johnson, J.A., Marcy, G.W., Fischer, D.A., Wright, J.T., Reffert, S., Kregenow, J.M., Williams, P.K.G., and Peek, K.M.G. 2007b, arXiv0711.4367J
\bibitem[kj]{kj} Kjeldsen, H. $\&$ Bedding, T.R. 1995, 293, 87
\bibitem[1997]{Kur97} K\"urster, M., Schmitt, J.H.M.M., Cutispoto, G. \& Dennerl, K. 1997, A\&A, 320, 831
\bibitem[lau97]{lau97} Laughlin, G., and Adams, F.C. 1997, ApJ, 491, 51
\bibitem[lin]{lin} Lin, D.N.C., Bodenheimer, P., and Richardson, D.C. 1996, Natur, 380, 606
\bibitem[mc]{mc} McArthur, B., Jefferys, W., and McCartney, J. 1994, AAS, 184, 2804
\bibitem[mu]{mu} Murray, N., Hansen, B., Holman, M., and Tremaine, S. 1998, Sci, 279, 69
\bibitem[nie]{nie} Niedzielski, A., Konacki, M., Wolszczan, A., Nowak, G., Maciejewski, G., Gelino, C.R., Shao, M., Shetrone, M., and Ramsey, L.W. 2007, ApJ, 669, 1354
\bibitem[pas]{pas} Pasquini, L., and Pallavicini, R. 1991, A$\&$A, 251, 199
\bibitem[pasq]{pasq} Pasquini, L., D\"{o}llinger, M. P., Weiss, A., Girardi, L., Chavero, C., Hatzes, A. P., da Silva, L., Setiawan, J. 2007, A$\&$A, 473, 979
\bibitem[pol]{pol} Pollack, J.B., Hubickyj, O., Bodenheimer, P., Lissauer, J.J., Podolak, M., and Greenzweig, Y. 1996, Icarus, 124, 62
\bibitem[qu]{qu} Quillen, A.C., and Holman, M. 2000, AJ, 119, 397
\bibitem[ra]{ra} Rasio, F.A., and Ford, E.B. 1996, Sci 274, 954
\bibitem[ref06]{ref06} Reffert, S., Quirrenbach, A., Mitchell, D.S., Albrecht, S., Hekker, S., Fischer, D.A., Marcy, G.W., and Butler, R.P. 2006, ApJ, 652, 661
\bibitem[san04]{san04} Santos, N.C., Israelian, G., and Mayor, M. 2004, A$\&$A, 415, 1153
\bibitem[sa03]{sa03} Sato, B., Ando, H., and Kambe, E. 2003, ApJ 597, L157
\bibitem[sa07]{sa07} Sato, B., Izumiura, H., Toyota, E., Kambe, E., Takeda, Y., Masuda, S., Omiya, M., Murata, D., Itoh, Y., Ando, H., and 4 coauthors 2007, ApJ, 661, 527
\bibitem[sa08]{sa08} Sato, B., Izumiura, H., Toyota, E., Kambe, Ikoma, M., Omiya, M., Masuda, S., Takeda, Y., Murata, D., Itoh, Y. 2008, arXiv0802.2590
\bibitem[sc]{sc} Scargle, J.D. 1982, ApJ, 263, 835
\bibitem[set03a]{set03a} Setiawan, J., Hatzes, A.P., von der L\"{u}he, O., Pasquini, L., Naef, D., da Silva, L., Udry, S., Queloz, D., and Girardi, L. 2003a, A$\&$A, 397, 1151
\bibitem[set03b]{set03b} Setiawan, J., Pasquini, L., da Silva, L., von der L\"{u}he, O., and Hatzes, A. 2003b, A$\&$A, 398, L19
\bibitem[set05]{set05} Setiawan, J., Rodmann, J., da Silva, L., Hatzes, A.P., Pasquini, L., von der L\"{u}he, O., de Medeiros, J.R., D\"{o}llinger, M.P., and Girardi, L. 2005, A$\&$A, 437, L31
\bibitem[schu]{schu} Schuler, S.C., Kim, J.H., Tinker, M.C., Jr., King, J.R., Hatzes, A.P., and Guenther, E.W. 2005, ApJ, 632, 131
\bibitem[val]{val} Valenti, J.A., Butler, R.P., and Marcy, G.W. 1995, PASP, 107, 966
\bibitem[van05]{van05} van Leeuwen, F. $\&$ Fantino, E. 2005, A $\&A$ 438, 791
\bibitem[va]{} van Leeuwen, F. 2007, $\&A$ 474, 653
\end{thebibliography}
\end{document}